\documentclass[10pt,a4paper]{article}
\usepackage{amsmath, amssymb}
\usepackage{graphicx}
\usepackage{subfig}
\usepackage{siunitx}
\usepackage{xcolor}
\usepackage[margin=2cm]{geometry}
\usepackage{cite}
\usepackage{authblk}

\title{Material Decomposition using Spectral Propagation-based Phase Contrast X-ray Imaging}
\author[1]{Florian Schaff}
\author[1]{Kaye~S.~Morgan}
\author[1]{James~A.~Pollock}
\author[1]{Linda~C.~P.~Croton}
\author[2]{Stuart~B.~Hooper}
\author[1]{Marcus~J.~Kitchen}
\affil[1]{School of Physics and Astronomy, Monash University, VIC 3800, Australia}
\affil[2]{The Ritchie Centre, Monash Institute of Medical Research, Monash University, VIC 3800, Australia}

\date{} 
\begin{document}

\maketitle

\begin{abstract}
Material decomposition in X-ray imaging uses the energy-dependence of attenuation to virtually decompose an object into specific constituent materials. X-ray phase contrast imaging is a developing technique that can enhance image contrast seen from weakly attenuating objects. In this paper, we combine spectral phase contrast imaging with material decomposition to both better visualise weakly attenuating features and separate them from overlying objects in radiography. We derive an algorithm that performs both tasks simultaneously and verify it against numerical simulations and experimental measurements of ideal two-component samples composed of pure aluminium and poly(methyl methacrylate). Additionally, we showcase first imaging results of a rabbit kitten's lung. The attenuation signal of a thorax, in particular, is dominated by the strongly attenuating bones of the ribcage, which combined with the weak soft tissue signal make it difficult to visualise the fine anatomical structures across the whole lung. In all cases, clean material decomposition was achieved, without residual phase contrast effects, from which we generate an un-obstructed image of the lung, free of bones. Spectral propagation-based phase contrast imaging has the potential to be a valuable tool, not only in future lung research, but also in other systems for which phase contrast imaging in combination with material decomposition proves to be advantageous.
\end{abstract}

For many X-ray imaging applications, time, dose, a rapidly changing sample, or other constraints prevent the use of three-dimensional X-ray computed tomography. In these cases, X-ray radiography still provides access to internal structures, although at the cost of volumetric information. It uses X-rays passing through an object from a single view, and, hence, only projected, two-dimensional information is available. Different parts of an object along the X-ray beam direction therefore appear superimposed in a radiographic image. Parts of an object containing strongly attenuating materials are often prevalent in these images and may obstruct the view of subtle changes in less-attenuating parts of an object. A common example of this is the dominance of bone over soft tissue in images of biological specimens. Spectral radiography, also often termed dual-energy X-ray imaging, is a method aimed at overcoming this intrinsic limitation present in projection-based X-ray imaging by combining images taken with different illuminating X-ray spectra \cite{Lehmann1981a}. As the relative change in attenuation with energy differs between elements, spectral radiography can be used to virtually separate an object into its constituent materials, a process also known as material decomposition \cite{Lehmann1981a}. Applications of spectral radiography include the determination of bone-mineral density \cite{Sorenson1990, Dinten2001a, Herve2002}, material discrimination in industrial and security applications \cite{Letang2004a, Chen2007a, Gilbert2016}, and bone removal in lung imaging \cite{Vock2009a, Carnibella2012a}.

\begin{figure}[b]
	\centering
	\includegraphics[width=.5\linewidth]{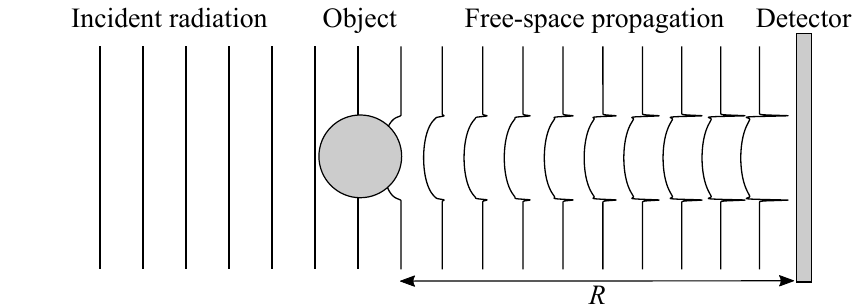}
	\caption[PBISetup]{Experimental realization of propagation-based X-ray imaging. An object is illuminated with a spatially coherent X-ray wave-field. Free space propagation is used to convert phase differences into intensity modulations which can be recorded by a detector positioned a distance $R$ downstream from the object.}
	\label{fig:PBISetup}
\end{figure}

During the last decades, various imaging modalities have emerged that, in addition to the attenuation, utilize the phase shifts that X-rays undergo when passing through an object. This includes propagation-based imaging (PBI) \cite{Snigirev1995, Cloetens1996}, using an analyzer crystal \cite{Rigon2007}, and various forms of imaging with a structured illumination, such as grating interferometry \cite{Momose2003, Weitkamp2005}, single-grid imaging \cite{Wen2010, Morgan2011}, edge-illumination \cite{Olivo2007}, and speckle tracking \cite{Berujon2012, Morgan2012}. A common recurring theme between them is the potential for increased contrast in weakly attenuating materials that are otherwise difficult to image with X-rays \cite{Davis1995, Momose1996}. The focus of this paper is on PBI, whose experimental realization is sketched in Fig. \ref{fig:PBISetup}. Illumination with sufficient spatial coherence passes through an object and is recorded some distance $R$ downstream of the object \cite{Cloetens1996}. In addition to the reduction of X-ray intensity due to attenuation, local shifts in the X-ray phase caused by the object manifest as measurable intensity variations after free space propagation. These so-called phase contrast fringes primarily appear in areas at the borders between different materials. As radiation used in PBI may be polychromatic \cite{Wilkins1996}, we seek to combine PBI with spectral imaging to achieve material decomposition with the added benefits of utilizing the phase information of X-rays.

Lung imaging is a common application of the aforementioned phase-contrast methods, given the relatively weak X-ray attenuation contrast of the prevalent air and soft tissue \cite{Kitchen2011, Schleede2012a, Kitchen2015, Garson2015, Hagen2019, Gradl2019}. PBI, in particular, benefits from the specific structure of many air-filled cavities within the lung that give rise to pronounced phase contrast effects, and has proven to be a valuable tool for pre-clinical lung imaging in recent years \cite{Hooper2007, Stahr2016, Bayat2018, Gradl2019a, Brooks2019}. The requirements of dynamic respiratory studies, which may be limited to two-dimensional imaging due to high temporal resolution \cite{Gradl2017}, cases of irreversible change or low dose requirements \cite{Hooper2007, Donnelley2013}, and the rise of pixelated spectral X-ray detectors in recent times have motivated the development of the method presented here.

\section{Spectral PBI}

In order to obtain quantitative information from PBI, which is required for material decomposition, so-called phase retrieval processing is applied to the acquired images. In general, phase retrieval using a single image can only be performed accurately by making additional assumptions (see Burvall \textit{et al.} \cite{Burvall2011} for a detailed overview). The most widely used phase retrieval algorithm, developed by Paganin \textit{et al.} \cite{Paganin2002a}, assumes an object to consist of a single material, which naturally can lead to artefacts in multi-material objects \cite{Beltran2010a, Burvall2013a, Haggmark2017a}. Other phase-retrieval algorithms relax the single-material assumption to some extent \cite{Wu2005a, Beltran2010a, Croton2018} or rely on more advanced iterative processing \cite{Hehn2018b}. In general, however, a single image is not sufficient for universal phase retrieval, and additional images under different experimental conditions have to be acquired, e.g. by changing the distance \cite{Gureyev1998a, Cloetens1999, Yu2018} and/or the X-ray energy \cite{Gureyev1998, Burvall2011}. Here, we focus on spectral imaging, and thus taking images at a constant propagation distance.

Several phase retrieval algorithms for spectral X-ray PBI have been proposed in the past. Gureyev \textit{et al.} \cite{Gureyev2001a, Gureyev2002} presented algorithms that use a material-independent scaling of attenuation and phase effects with respect to the X-ray energy. This model works at very low X-ray energies and in the absence of absorption edges for which photoelectric absorption is the primary source of attenuation and Compton scattering is negligible \cite{Gursoy2013}. This is important, given that Compton scattering is the predominant cause of attenuation for all elements at very high X-ray energies. Wu \textit{et al.} \cite{Wu2005a} used the realization that, at high X-ray energies, both the X-ray attenuation coefficient $\mu$ and the refractive index decrement $\delta$ (which describes the phase effects) are proportional to the electron density, allowing them to derive a phase retrieval algorithm very similar to the single material approximation of Paganin \textit{et al.} \cite{Paganin2002a}. 

\begin{figure}[tbp]
	\centering
	\includegraphics[width=.5\linewidth]{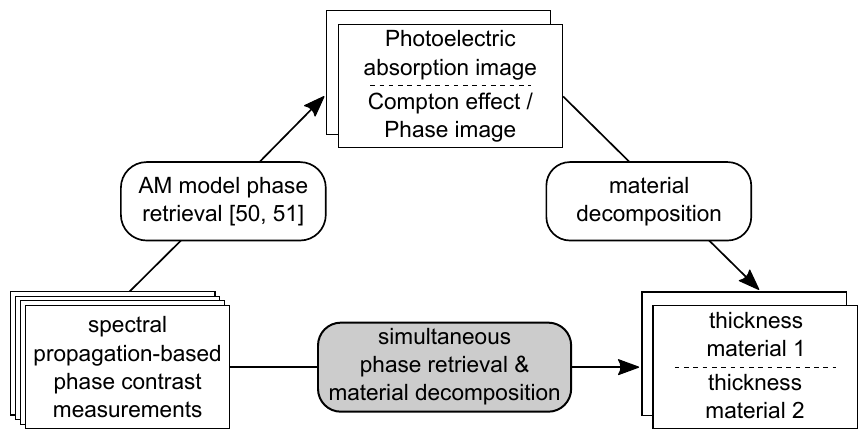}
	\caption[Flowchart]{Overview of the algorithms discussed in this paper. Starting from spectral PBI data, the algorithms based on the Alvarez-Macovski model yields phase-retrieved maps of the projected photoelectric absorption and Compton effect \cite{Gursoy2013, Li2018}. In a second step, this result can be decomposed into the projected thickness maps of two basis materials. We present an algorithm that performs phase retrieval and material decomposition simultaneously.}
	\label{fig:flowchart}
\end{figure}

If neither of the above conditions are met, algorithms developed by G\"ursoy \& Das \cite{Gursoy2013} and Li \textit{et al.}\cite{Li2018} that employ the Alvarez-Macovski (AM) model \cite{Alvarez1976a} to describe attenuation as a combination of photoelectric absorption and Compton scattering can be used for the intermediate energy range. The phase retrieved result of these algorithms represents maps of the local photoelectric absorption and Compton scattering magnitudes. With known ratios of these contributions for two materials in question, one could augment these algorithms with an additional basis transform that converts the result into thickness maps of the two materials in question. The relationship between the work presented in this paper and the aforementioned algorithms of \cite{Gursoy2013} and \cite{Li2018} is illustrated in Fig. \ref{fig:flowchart}. In the following, we show that material decomposition removes the need for the AM model, and we derive an algorithm that performs phase retrieval and material decomposition in a single step. This has potential benefits over using the AM model, which we discuss at the end of the paper.

\subsection*{Linearized Transport of Intensity Equation}

In this section we derive our proposed algorithm for simultaneous phase retrieval and material decomposition. Consider two-dimensional X-ray illumination in the $(x, y)-$plane, with intensity $I_{\textrm{in}}(E,x,y) \equiv I_{\textrm{in}}(E)$ and uniform phase passing through an object in beam direction, $z$. We set the exit plane of the object to be at $z=0$, and within the projection approximation \cite[p.~71]{Paganin2006}, the resulting phase, $\Phi(E,x,y,z=0) \equiv \Phi(E)$, and intensity, $I(E,x,y,z=0) \equiv I_{0}(E)$, distributions are given as:

\begin{equation}\label{eq:mu}
I_{0}(E) = I_{\textrm{in}}\exp{\left(-\int\mu(E)\right)}\textrm{d}z;~\Phi(E) = -\frac{2\pi}{\lambda}\int\delta(E)\textrm{d}z.
\end{equation}

Here, $\mu(E)$ is the linear attenuation coefficient, $\delta(E)$ the refractive index decrement, and $\lambda$ the X-ray wavelength. As all integrals involved are along the $z$-direction, we omit $\textrm{d}z$ in the following notation. Note that we explicitly stated the dependency of these equations on the X-ray energy, $E$, which plays a pivotal role in spectral imaging. Given $\Phi(E)$ and $I_{0}(E)$, we can use the transport of intensity equation (TIE) to describe the spatial evolution of this wave-field for propagation in $z$. Under the finite difference approximation, the TIE relates the propagated intensity $I(E,x,y,z=R) \equiv I_{R}(E)$ at distance $R$ to $\Phi(E)$ and $I_{0}(E)$ as \cite{Gureyev2002}:

\begin{equation}\label{eq:TIE}
I_{R}(E) = I_{0}(E) - I_{0}(E)\frac{R\lambda}{2\pi}\nabla^2\Phi(E) -\frac{R\lambda}{2\pi}\nabla I_{0}(E) \nabla\Phi(E).
\end{equation}

Without large intensity gradients in $I_{0}(E)$, we can disregard the last term \cite[p.~297]{Paganin2006} and insert eq. \ref{eq:mu} into eq. \ref{eq:TIE}:

\begin{equation}
I(E) \equiv \frac{I_{R}(E)}{I_{\textrm{in}}(E)} = \exp{\left(-\int\mu(E)\right)}\left(1 + R\nabla^2\int\delta(E)\right)
\end{equation}

As a last step, we take the logarithm, assume a moderate magnitude of phase effects to use the approximation $\ln(1+x)=x$ for $x\ll1$ \cite{Gureyev2001a}, and arrive at the linearised TIE that also serves as a starting point for Gursoy \& Das \cite{Gursoy2013}:

\begin{equation}\label{eq:TIElin}
-\ln\left[I(E)\right] = \int\mu(E) - R\nabla^2\int\delta(E).
\end{equation}

\subsection*{Phase retrieval algorithm}
Phase retrieval aims to obtain $\int\mu (E)$ and $\int\delta (E)$, given one or more intensity measurements, $I(E)$. Taking the Fourier transform, $\mathcal{F}$, of eq. \ref{eq:TIElin} allows us to use the Fourier differentiation theorem and express the Laplacian operator in terms of the two-dimensional Fourier components as $\nabla^2=-\mathbf{k}^2$:

\begin{equation}\label{eq:FTIElin}
-\mathcal{F}\left[\ln\left(I(E)\right)\right] = \mathcal{F}\left[\int\mu(E)\right] + R\mathbf{k}^2\mathcal{F}\left[\int\delta(E)\right].
\end{equation}

It is evident that, without \textit{a priori} information about the imaged object, $n \geq 2$ independent measurements are required to retrieve the two unknowns, $\int\mu(E)$ and $\int\delta(E)$. However, this is in general not possible, even with multiple measurements. Taking a closer look at eq. \ref{eq:FTIElin} reveals that any system of equations is inherently unstable at $\mathbf{k}=0$, as $\delta(E)$ vanishes completely. Enforcing some proportionality between $\int\mu(E)$ and $\int\delta(E)$ provides the necessary stability at $\mathbf{k}=0$, but implies the well-known single material approximation \cite{Paganin2002a} for $E = const$. Thus, we are required to use spectral information for samples consisting of multiple materials. Given the energy dependence of $\mu(E)$ and $\delta(E)$, changing the X-ray energy adds more unknowns, and we retain an under-determined system. Even though analytically $\delta(E) \propto E^{-2}$ \cite[p.~71]{Als-Nielsen2011}, the energy dependence of $\mu(E)$ is more complex and differs for all elements. This is what makes material decomposition a possibility in attenuation imaging, but prevents us from formulating a general energy dependence for $\mu(E)$. All of this considered, it is clear that, in addition to some proportionality between $\int\mu(E)$ an $\int\delta(E)$, we also require knowledge about how $\int\mu(E)$ changes with energy.

The previously mentioned algorithms by G\"ursoy \& Das \cite{Gursoy2013} and Li \textit{et al.} \cite{Li2018} use the AM model to approximate the energy dependence of $\mu(E)$ as a combination of two basis functions that represent photoelectric absorption and Compton scattering, respectively \cite{Gursoy2013, Li2018}. As it turns out, the Compton part mathematically shares the same linear dependence on the overall electron density, $\rho_{\textrm{e}}$, as $\delta$, which provides the required coupling between $\mu(E)$ and $\delta(E)$ and thus algorithmic stability at $\mathbf{k}=0$.

Here, we pursue a different approach. In light of our goal of material decomposition, a set of basis functions defined by the specific materials in question suggests itself. This approach has been used for material decomposition since the early works on spectral attenuation X-ray imaging \cite{Lehmann1981a} and common applications are e.g. bone mineral density determination \cite{Hofmann2016}, or detection of calcifications in mammography \cite{Kappadath2003}. In an object composed of $m$ basis materials, with projected thicknesses $t_{j=1,..,m}$, $\int\mu(E)$ and $\int\delta(E)$ are determined using a linear combination of the respective tabulated $\mu_j(E)$ and $\delta_j(E)$ values as: 

\begin{equation}\label{eq:musum}
\int\mu(E) = \sum_{j}^{m}t_j\mu_j(E);~\int\delta(E) = \sum_{j}^{m}t_j\delta_j(E) 
\end{equation}

Not only does eq. \ref{eq:musum} define the energy dependence of $\int\mu(E)$ and $\int\delta(E)$, it provides an intrinsic coupling between the two through $t_j$ for each of the materials. Given $n$ intensity measurements, $I_{i=1,..,n;~n \geq m}$, taken with discrete X-ray energies, $E_i$, we can substitute eq. \ref{eq:musum} into eq. \ref{eq:FTIElin} and formulate a linear system of equations:

\begin{equation}\label{eq:2matmodel}
-\mathcal{F}\left(\ln I_\textrm{i}\right) =  \sum_{j}^{m}\underbrace{\left[\mu_j(E_\textrm{i}) + R\mathbf{k}^2\delta_j(E_i)\right]}_{\let\scriptstyle\textstyle \substack{\mathbf{A}_{ij}}}\mathcal{F}\left(t_j\right)
\end{equation}

\begin{equation}\label{eq:matrix2matTIE}
\underbrace{\begin{bmatrix} 
	\mathbf{A}_{11} & \mathbf{A}_{12} & \hdotsfor{1} & \mathbf{A}_{1m} \\ 
	\mathbf{A}_{21} & \mathbf{A}_{22} & \hdotsfor{1} & \mathbf{A}_{2m} \\ 
	\vdots & \vdots & \ddots & \vdots \\ 
	\mathbf{A}_{n1} & \mathbf{A}_{n2} & \hdotsfor{1} &  \mathbf{A}_{nm}
	\end{bmatrix}}_{\let\scriptstyle\textstyle \substack{\mathbf{A}}}
\underbrace{\begin{bmatrix} 
	\mathcal{F}(t_{\textrm{1}})\\ 
	\mathcal{F}(t_{\textrm{2}})\\
	\vdots \\
	\mathcal{F}(t_{\textrm{m}})\\  
	\end{bmatrix}}_{\let\scriptstyle\textstyle \substack{\mathbf{t}}}
= 
\underbrace{\begin{bmatrix} 
	-\mathcal{F}(\ln{I}_{\textrm{1}})\\ 
	-\mathcal{F}(\ln{I}_{\textrm{2}})\\ 
	\vdots\\ 
	-\mathcal{F}(\ln{I}_{\textrm{n}})
	\end{bmatrix}}_{\let\scriptstyle\textstyle \substack{\mathbf{I}}}
\end{equation}

Note that $\mathbf{A}_{ij}$, $t_{j}$, and $I_i$ are all spatially resolved, two-dimensional quantities in $(x, y)$, and the element-wise Hadamard product is to be used for multiplications, where appropriate. We obtain the least squares solution of eq. \ref{eq:matrix2matTIE} by solving the associated normal equation $\textbf{A}^\textrm{T}\textbf{A}\textbf{t} = \textbf{A}^\textrm{T}\textbf{I}$, a similar approach to \cite{Gursoy2013}. As a typical object consists of a limited number of materials $m$, and in many cases additional materials can be expressed as a linear combination of the first two \cite{Lehmann1981a}, only a small $m \times m$ matrix needs to be inverted for every entry in $\textbf{A}^\textrm{T}\textbf{A}$. Computation of $\textbf{t} = (\textbf{A}^\textrm{T}\textbf{A})^{-1}\textbf{A}^\textrm{T}\textbf{I}$ therefore can be performed in seconds using modern computer hardware. Additionally, the calculation of $(\textbf{A}^\textrm{T}\textbf{A})^{-1}\textbf{A}^\textrm{T}$ only has to be performed once and can be used for subsequent sets of input images $\textbf{I}$ with the same experimental parameters, e.g. during a CT scan or temporal study, which is computationally vastly more efficient than iterative methods. As a last step, the inverse Fourier transform $\mathcal{F}^{-1}$ is used to obtain the material decomposed, phase retrieved thickness maps $t_j$ from $\textbf{t}$.

\section{Results}

\subsection*{Simulations}

\begin{figure*}[tbhp]
	\centering
	\includegraphics[width=1\linewidth]{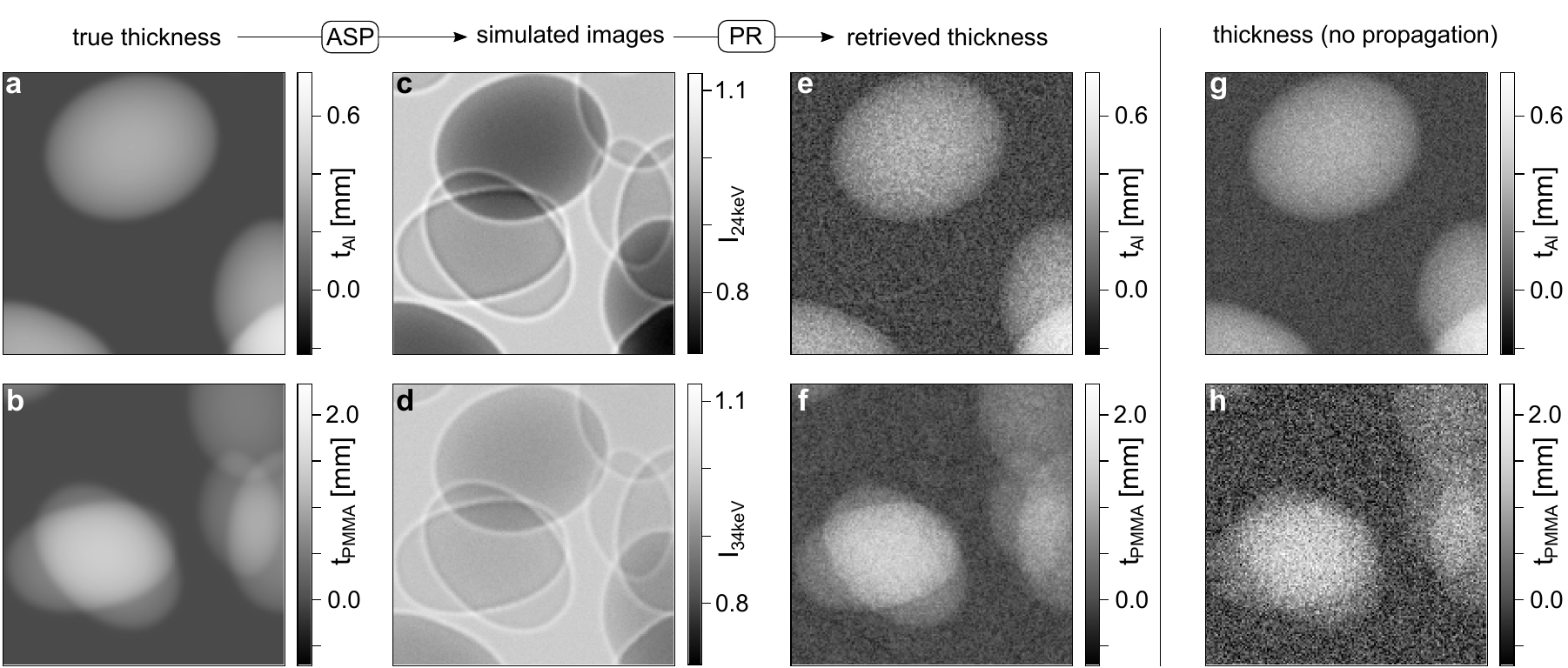}
	\caption[Simulation of basis material phase retrieval]{Simulation of basis material phase retrieval. (a) and (b): Input thickness maps for Al and PMMA. (c) and (d): Angular spectrum propagated (ASP) intensity at $R=\SI{20}{\centi\meter}$ downstream of the sample for monochromatic \SI{24}{\kilo\electronvolt} and \SI{34}{\kilo\electronvolt} X-ray illumination with $0.4\%$ Poisson-noise. (e) and (f): Phase retrieved thickness maps for Al and PMMA from propagated intensity measurements. (g) and (h): Retrieved thickness maps for Al and PMMA with setting $R=\SI{0}{\meter}$ in eq. \ref{eq:2matmodel}}.
	\label{fig:BasisMaterialsSimulation}
\end{figure*}

We first tested the algorithm derived in the previous section under controlled conditions, without potential experimental uncertainties, by using synthetic data of a numerically simulated phantom. The results are depicted in Fig. \ref{fig:BasisMaterialsSimulation}. Panels (a) and (b) show the two-dimensional projected input thickness maps of aluminium (Al) and poly(methyl methacrylate) (PMMA) that were used to model an ideal, two-component object under the assumptions of parallel illumination and the projection approximation \cite{Paganin2006}. From this, we generated synthetic PBI data with a propagation distance $R=\SI{20}{\centi\meter}$ and detector pixel size of \SI{6.5}{\micro\meter} for monochromatic illumination of $E=\SI{24}{\kilo\electronvolt}$ and $E=\SI{34}{\kilo\electronvolt}$. All simulation parameters were chosen to closely match those of the experimental data, presented in the next section. We created the input wave-fields in the exit plane of the sample, given by $I_0(E)$ and $\Phi(E)$, using theoretical $\mu(E)$ and $\delta(E)$ values for Al and PMMA from the xraylib library for python \cite{Schoonjans2011}. These were then up-sampled by a factor of five to increase accuracy, and the angular spectrum propagator (ASP) \cite{Paganin2006} was used to obtain the propagated intensity distributions. Lastly, Poisson-noise with a magnitude of $0.4\%$ in the background region, similar to our experimental data, was applied to arrive at the images shown in Fig. \ref{fig:BasisMaterialsSimulation} (c) and (d) for $E=\SI{24}{\kilo\electronvolt}$ and $E=\SI{34}{\kilo\electronvolt}$, respectively. These images were then used as an input to our proposed algorithm.
The resulting phase retrieved thickness maps $t_\textrm{Al}$ and $t_\textrm{PMMA}$ are given in panels (e) and (f). Compare this result with the ground truth given in (a) and (b). Quantitative thickness information is recovered for both materials, and the strong phase-contrast fringes are nearly completely removed. Furthermore, it is clear to see that the resulting images are visibly more noisy than the input data. From conventional spectral X-ray radiography, it is well known that material decomposition typically increases noise in the final images \cite{Lemacks2002a, Herve2002, Letang2004a, Vock2009a}. In the absence of phase retrieval, this result is therefore to be expected. However, phase retrieval, in particular with the single-material approximation \cite{Paganin2002a}, has the potential to significantly reduce image noise \cite{Gureyev2017a, Kitchen2017}. Since our algorithm performs a combination of phase retrieval and material decomposition, which have opposite effects on image noise, simultaneously, it is important to compare these results to those from conventional material decomposition without phase retrieval. For this purpose, we repeated the simulation under identical conditions, except with a propagation distance $R=\SI{0}{\meter}$. The resultant images are displayed in panels (g) and (h). From this, we can clearly see that simultaneous phase retrieval and material decomposition on PBI data appears to reduce noise in $t_\textrm{PMMA}$ at the expense of increased noise in $t_\textrm{Al}$, compared to the case without phase contrast. Quantitatively, propagation caused a decrease of the root-mean-square error ($rms$) between $t_\textrm{PMMA}$ and $t_\textrm{PMMA,input}$ by a factor of approximately 3.15 and a simultaneous increase of $rms(t_\textrm{Al}-t_\textrm{Al,input})$ by a factor of 1.5. Given the typically worse attenuation contrast of weakly absorbing materials, a trade-off in noise in favour of the weakly absorbing material, as is the case for PMMA here, is desirable. A comprehensive study of the exact noise characteristics of our algorithm is beyond the scope of this paper and will be addressed in future work.

\subsection*{Experimental Data}

\begin{figure*}[tbhp]
	\centering
	\includegraphics[width=1\linewidth]{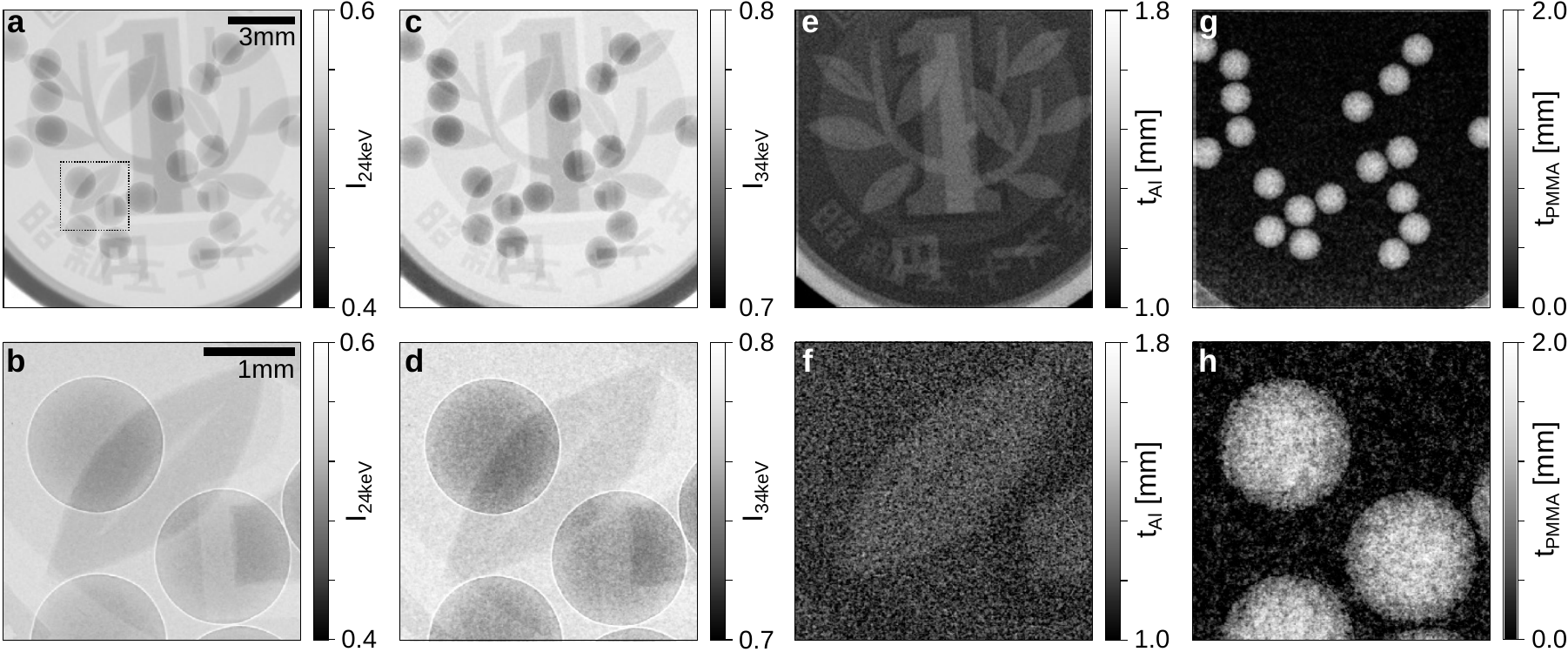}
	\caption[YenPMMA]{Propagation-based X-ray images of a test sample consisting of a 1 yen coin, made from pure aluminium, and several PMMA spheres at (a) 24 keV (c) 34 keV. Note that these figures are displayed with different intensity ranges. Spectral phase retrieval and material decomposition separates the images into thickness maps for (e) aluminium, and (g) PMMA with no residual phase contrast visible. (b), (d), (f) and (h): Magnified views of the area marked in (a) for the images shown in the upper row.}
	\label{fig:YenPMMA}
\end{figure*}

Spectral PBI experiments were performed at the Medical and Imaging I Beamline, BL20B2, of the SPring-8 Synchrotron in Hy$\overline{\mbox{o}}$go, Japan \cite{Goto2001}. At this beamline, a large distance from the bending magnet source of over \SI{200}{\meter}, paired with a double-crystal monochromator \cite{Yabashi2003} offers tunable monochromatic illumination with sufficient spatial coherence over an extended field of view, ideal for our PBI experiments. Images were recorded using a \SI{6.5}{\micro\meter} pixel size Hamamatsu C11440-52U series sCMOS camera equipped with a \SI{10}{\micro\meter} thick Gadox ($\textrm{Gd}_2\textrm{O}_2\textrm{S}: \textrm{Tb}^{2+}$; P43) scintillator coupled to the sensor via a straight fibre optic. 

Fig. \ref{fig:YenPMMA} shows the experimental results of a composite object consisting of a 1981 Japanese 1 yen coin and several PMMA spheres of approximately \SI{1.5}{\milli\meter} diameter. The coin is made of pure aluminium with a ridge height of \SI{1.5}{\milli\meter}, which in combination with the PMMA spheres creates an ideal two-component sample to test our algorithm. Magnified images of the rectangular area indicated in (a) are given in the lower row for the respective full frames seen directly above. Fig. \ref{fig:YenPMMA} (a) and (c) shows spectral PBI data at a constant propagation distance $R=\SI{0.2}{\meter}$, with X-ray energies of $E_1 = \SI{24}{\kilo\electronvolt}$ and $E_2 = \SI{34}{\kilo\electronvolt}$, respectively.
Owing to the penetrating nature of X-rays, both sides of the coin, as well as the PMMA spheres, are superimposed in these images. In addition, strong phase contrast fringes along the boundaries of the PMMA spheres are clearly visible alongside the conventional absorption contrast. The results of simultaneous phase retrieval and material decomposition, namely $t_\textrm{Al}$ and $t_\textrm{PMMA}$, are given in panels (e) and (g). As expected, the algorithm is able to perform phase retrieval and decompose the spectral PBI data into $t_\textrm{Al}$ and $t_\textrm{PMMA}$. Both components are well separated, and there is no visible cross-contamination between $t_\textrm{Al}$ and $t_\textrm{PMMA}$. Quantitatively, the expected maximum thickness through the center of the PMMA spheres and along the ridge of the coin matches the expected values of \SI{1.5}{\milli\meter} well. We also note that $t_\textrm{Al}$ and $t_\textrm{PMMA}$ are visibly more noisy than the raw data, as expected from the simulations. Nonetheless, the beneficial effect of phase retrieval on $t_\textrm{PMMA}$ when paired with $t_\textrm{Al}$ is an inherent characteristic of the algorithm, and thus is independent of the sample.

\begin{figure*}[tbhp]
	\centering
	\includegraphics[width=1\linewidth]{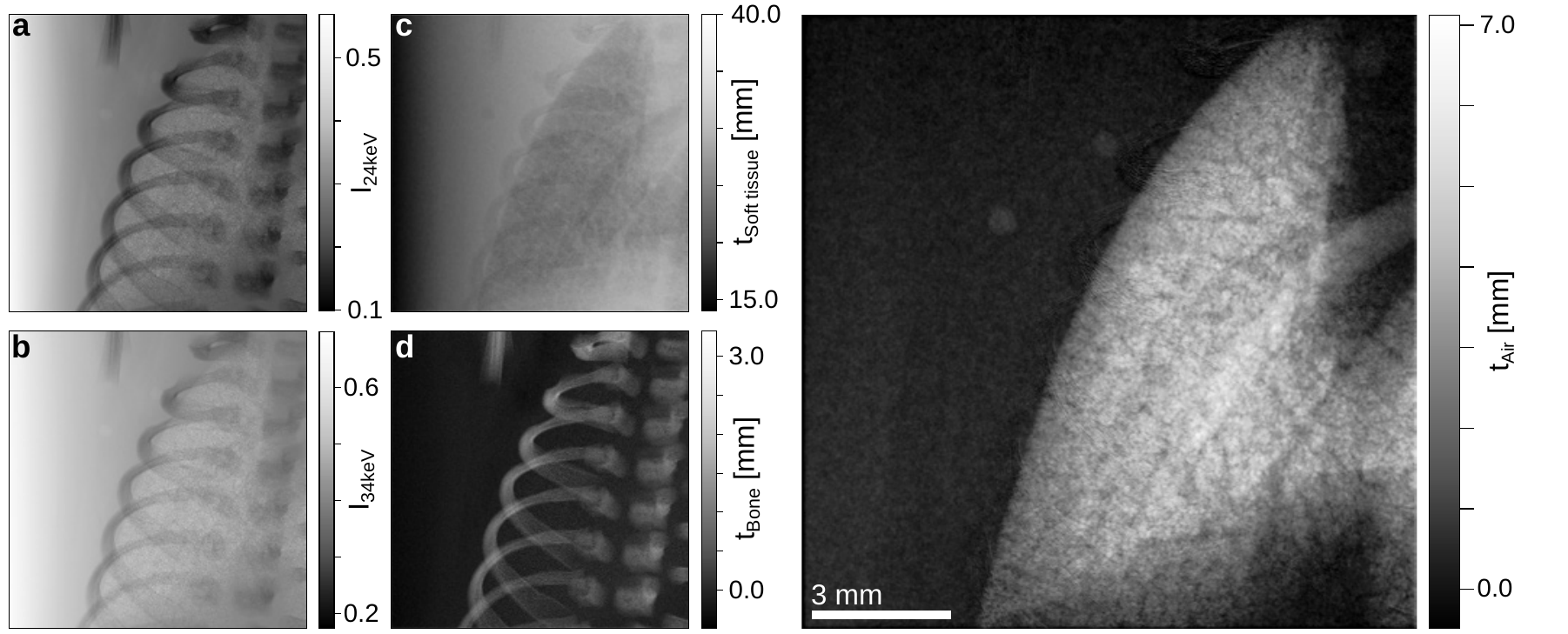}
	\caption[RKLittleFlash]{Propagation-based X-ray images of the lung area of a deceased neonatal rabbit kitten embedded in agarose gel at (a) $E=\SI{24}{\kilo\electronvolt}$ and (b) $E=\SI{34}{\kilo\electronvolt}$. After spectral phase retrieval and material decomposition, thickness maps of the (c) soft tissue (d) bone are obtained. (e) Image of the projected lung air volume, noting that the image processing steps required to extract this air image are explained in the main text.}
	\label{fig:RKLittleFlash}
\end{figure*}

Finally, we take a look at the scientific problem stated at the outset of this paper: spectral PBI of the lung. Results of a deceased neonatal rabbit kitten are presented in Fig \ref{fig:RKLittleFlash}. As we needed to take spectral images sequentially, the animal was fixed in agarose gel to prevent deformation during the measurement. Radiographic images of the chest area containing one side of the lung were again taken at $E=\SI{24}{\kilo\electronvolt}$, shown in (a), and $E=\SI{34}{\kilo\electronvolt}$, shown in (b). The chest consists primarily of two materials (besides air): soft tissue and bone, which are superimposed in the images. Naturally, not all soft tissues are the same, and the object also includes a large amount of agarose gel and the plastic sample container. However, the X-ray optical properties of these materials are approximately equal to those of soft tissue, and were considered as equal. For phase retrieval, we therefore used the theoretical values for cortical bone and soft tissue as defined by the International Commission on Radiological Protection (ICRP) \cite{Schoonjans2011}. Given that bone and soft tissue have very comparable X-ray optical properties to Al and PMMA, and our experiments were performed at the same X-ray energies, the observation that phase retrieval enhances the image of the weakly attenuating constituent can be translated to this system. 
After applying phase retrieval and material decomposition, we obtained thickness maps for soft tissue, $t_\textrm{Soft tissue}$, and bone, $t_\textrm{Bone}$, which are presented in Fig \ref{fig:RKLittleFlash} (c) and (d), respectively. The bone thickness image illustrates excellent material decomposition - all of the relevant bone structures, which includes parts of the ribcage, spine, and one limb, are well isolated from the soft tissues. Nonetheless, bone structures can clearly be identified in the soft tissue image. This is not caused by a residual bone signal but rather the lack thereof.  By computationally separating soft tissue and bone, removal of bone leaves behind voids in soft tissue, which can be seen here. As we are primarily interested in the lung, we can virtually fill these voids by calculating the sum $t_\textrm{Soft tissue} + t_\textrm{Bone}$. Within our approximations, this represents the aggregate thickness of all materials that are not air, i.e. everything except for the lung air volume. The projected thickness of a cylindrical container follows a function of the form $a\sqrt{b^2 - (x-c)^2}$. We fitted this function in a region outside the lung and obtained a smoothly varying map of the background. Any deviations from this are consequently caused by the lung, and we subtracted $t_\textrm{Soft tissue} + t_\textrm{Bone}$ from the fitted background to arrive at the positive projected air thickness within the lung, $t_\textrm{Air}$, depicted in panel (e). This gives us an isolated image of the projected lung air volume, free of bones, as was the goal of this paper. The trachea and fine-scale fluctuations in thickness due to the branch-like structure of the lungs are clearly visible. The two small, circular structures next to the lung are air bubbles formed in the agarose gel during sample preparation and, consequently, remain in the air-only image.

We shall note that the last image processing step relies on knowledge of the total object thickness, which we found here by using a container with a simple geometry. This step would not be required if a square container were used. When experimental requirements, e.g. during \textit{in-vivo} experiments, prevent the use of such a well-known container, this is a non-trivial task; however, laser profilometry has been suggested as a way to obtain the total thickness of an object in this case \cite{Myers2008}.

\section{Discussion}

The results shown here demonstrate that phase retrieval with simultaneous material decomposition is possible using spectral PBI data. In contrast to existing methods \cite{Gursoy2013, Li2018}, the phase retrieval step relies on the material coefficients, rather than the Alvarez-Macovski model. To realize the advantage of this, one needs to consider the limitations of the AM model. The Klein-Nishina formula used in the AM model describes Compton scattering from free, rather than bound, electrons. Consequently, when the electron binding energy is large, e.g. for low X-ray energies and high Z materials, the AM model may fail to provide an accurate description for attenuation. We shall, however, note that the AM model provides a good basis to describe attenuation within its limitations. Biologic imaging, in particular, is typically within these limits, given that the X-ray energies commonly used are much larger than the electron binding energies of the materials involved. Considering that the attenuation and phase coefficients are required for material decomposition in the first place, there is no need for the AM model and the risk of introducing errors. Another potential benefit lies in the fact that, on its own, the AM model cannot describe attenuation in the presence of absorption edges without the use of additional specified basis materials, which has been demonstrated experimentally for X-ray attenuation CT \cite{Schlomka2008a}. As our algorithm inherently relies on specific basis materials, absorption edges can easily be incorporated. 

Due to the use of material coefficients, the most obvious limitation of the proposed algorithm is the need for a priori knowledge of the constituent materials of an object. This, however, is a necessary requirement for material decomposition already and would also be needed to perform AM-based material decomposition. Nonetheless, varying material composition, or, in the context of biological imaging, physiological deviations can be a potential source of inaccuracy. This is a general problem in the field of material decomposition, and alternate algorithms would suffer from similar limitations.

Our experimental results further indicate how PBI, in the context of material decomposition, is able to reduce image noise of the weakly attenuating component in our systems  studied. Given that phase contrast is primarily used to enhance the visibility of weakly attenuating materials, this underlines the potential benefit of performing spectral PBI for material decomposition. During our experiments, we chose $E_1$ to match the typical energy used in similar lung imaging studies \cite{Kitchen2015}, and $E_2$ was a compromise between spectral separation and signal strength. The optimal parameters vary for each specific system \cite{Herve2002, Letang2004a}, and future research aimed at optimizing the experimental parameters for spectral PBI of the lungs of small animals needs to be undertaken.

We end this paper by noting that, for this proof of concept study, we acquired images at different energies sequentially during the experiments. It is clear that simultaneous acquisition is preferable for several reasons, e.g. the possibility to perform dynamic spectral imaging, or to avoid potential misalignment between the scans, a difficulty that is intrinsically associated with multiple-distance methods. Out of various approaches in attenuation-based X-ray imaging, the use of energy-resolving X-ray detectors has also been demonstrated for spectral PBI \cite{Vazquez2019}. Given the ongoing development of such detectors, we expect these devices to be widely used in the future, which will greatly aid the availability of time-resolved spectral X-ray data. We plan to use the presented method in quantitative respiratory studies on small animal models in the future.

\section*{Acknowledgment}
We thank Prof. David Paganin and Dr. Thomas Li for fruitful discussions. The synchrotron radiation experiments were performed at beamline BL20B2 of SPring-8 with the approval of the Japan Synchrotron Radiation Research Institute (JASRI) (Proposal No. 2019A0150). We are grateful for the support provided by the BL20B2 beamline staff. Funding for this research was provided by the Australian Research Council's Discovery Grant (DP170103678) and Future Fellowship schemes (MJK, FT160100454; KSM, FT180100374); an NHMRC Project Grant (1102564); a Veski Victorian Postdoctoral Research Fellowship (VPRF); the German Excellence Initiative; and the European Union Seventh Framework Program (291763). JP has a RTP (Research Training Program) scholarship and a J.~L.~Williams top up scholarship. We acknowledge travel funding provided by the International Synchrotron Access Program (ISAP) managed by the Australian Synchrotron, part of ANSTO, and funded by the Australian Government (AS/IA173/14218). Our phase retrieval code is available at doi.org/10.26180/5d9d7a41c0ab7.

\section*{Ethics Statement}
The deceased rabbit kitten was scavenged from preceding terminal animal imaging experiments that had approval from the SPring-8 and Monash University animal ethics committees.
%

\end{document}